\documentstyle[PASJadd]{PASJ95}
\newcommand{\vol}{49}
\newcommand{\no}{5}
\newcommand{\lett}{}
\newcommand{\spage}{}
\newcommand{\rdate}{1997 February 28}
\newcommand{\adate}{1997 August 16}

\newcommand {\beq}{\begin{equation}}
\newcommand {\eeq}{\end{equation}}

\newcommand {\vk}{$\rm km\; s^{-1}\;$}
\begin{document}
\title{On the Keplerian Rotation Curves of Galaxies}
\author{Mareki {\sc Honma} and Yoshiaki {\sc Sofue}\\
{\it Institute of Astronomy, Faculty of Science, The University of Tokyo, Mitaka, Tokyo, 181}\\
{\it E-mail : honma@milano.mtk.nao.ac.jp}}
\abst{We propose a criterion for examining whether or not the uncertainty of the outer rotation curves is sufficiently small to distinguish a Keplerian rotation curve from flat ones.
We have applied this criterion to both Keplerian and non-Keplerian rotation curves so far obtained, and investigated their relative fraction.
We also studied the minimum extent of the dark halos indicated by non-Keplerian rotation curves using the criterion.
We have found that one cannot rule out the possibility that a significant fraction of rotation curves become Keplerian within 10-times the disk scale length.
If the Keplerian rotation curves so far observed trace the mass truncation, several galaxies may have rather small halos, the extent of which is not larger than twice that of the optical disk.}
\kword{Dark Matter --- Galaxies: spiral --- Rotation}
\maketitle
\thispagestyle{headings}
%
% section 1
%
\section{Introduction}

A rotation curve is one of the most powerful tools for studying the mass distribution in galaxies.
Rotation curves have been extensively studied by the optical spectroscopy of {\sc Hii} lines, by radio observations of {\sc Hi} and molecular gases, and by their combinations (e.g., Bosma 1981; Rubin et al. 1985; Mathewson et al. 1992; Sofue 1996, 1997).
In most cases, a dark halo is required to reproduce the rotation curve beyond the optical disk (e.g., Carignan, Freeman 1985; Sancisi, van Albada 1987).

While rotation curves are referred to as {`}flat{'} as a first-order approximation, they are not completely featureless.
Rubin et al. (1985) argued that the profile of the rotation curves is dependent on the luminosity, which is supported by later studies (e.g., Persic, Salucci 1991; Persic et al. 1996).
There are also a number of rotation curves which show a velocity decrease at the optical edge (e.g., Salucci, Frenk 1989).
Salucci and Frenk (1989) argued that a rotation curve can decline outside a massive disk if the rotation velocity of the disk is larger than that of the extended halo.
The decrease in the rotation velocity is small, typically 10 -- $20$ \vk.

Recently, however, several declining rotation curves, which can be approximated by a Keplerian, have been found (e.g, Carignan, Puche 1990a; J\"ors\"ater, van Moorsel 1995; Olling 1996; Jore et al.1996).
The Milky Way Galaxy may also have such a rotation curve (Honma, Sofue 1996).
Although one cannot totally rule out mass distribution beyond the Keplerian region, the simplest interpretation of these rotation curves is mass truncation, including a dark halo.
Besides these declining rotation curves, there have been several indications of a possible truncation of the dark halo (e.g., Gottesman, Hunter 1982; Dubinski et al.1996; Bland-Hawthorn et al.1997).
If some dark halos are truncated at radii not far from the optical disk, this raises several questions:
how far the halo of a non-Keplerian galaxy extend; what is different between the halos of these galaxies; 
how much fraction of the galaxies have such small halos; and so on.
Since Keplerian rotation curve is slowly varying with the radius ($V\propto R^{1/2}$), and since any rotation curve rarely declines faster than that, it is important to consider the accuracy of the rotation curves for answering these questions.
For instance, suppose that the rotation curve of a galaxy is obtained out to $R_{\rm out}$.
If the real rotation curve becomes Keplerian beyond $R_{\rm dec}$, $R_{\rm out}$ must be larger than $R_{\rm dec}$ in order to detect any decline.
Even when this is satisfied, the rotation velocity should be determined with high accuracy compared to the decrease in the rotation velocity in the Keplerian region.
If we have only a few independent measurements of the rotation velocity between $R_{\rm dec}$ and $R_{\rm out}$, it is required that
\beq
2 \sigma \le V(R_{\rm dec})\left[1-\left(\frac{R_{\rm dec}}{R_{\rm out}}\right)^{1/2}\right]
\eeq
in order to discriminate Keplerian rotation curves from flat rotation curves.
Here, $\sigma$ is the size of the error bar (the velocity is expressed as $V\pm \sigma$), and the right-hand term expresses the velocity decrease in the Keplerian region.
If this condition is not satisfied, the rotation curve may be observed as a `flat' rotation curve, even in the Keplerian region.

In this paper we examine these conditions of the rotation curves so far obtained, and try to answer the questions raised above.
The plan of this paper is as follows:
In section 2 we describe the sample used in the present studied.
In section 3 we investigate how much the fraction of galaxies could have Keplerian rotation curves.
We discuss the extent of the dark halo in section 4.

%
% section 2
%
\section{The Sample}

In the present paper, we concentrate on the {\sc Hi} rotation curves of galaxies.
The rotation curves obtained by optical observations are not suitable for this study, because they are not likely to extend outside of the optical disk.
The rotation curves observed in CO lines are also unsuitable, since the molecular gas is usually more concentrated than the HI gas (Sofue et al.1995).
The sample galaxies consist of those which are observed in {\sc Hi}, and satisfy the following selection criteria:
1) galaxies later than Sa and not strongly distorted; 2) inclination larger than $45^\circ$, so that the effect of warping is small, 3) necessary data for the present analysis, such as the observational parameters and the optical profiles, are available.
There are 45 galaxies in the sample.
The basic data for the sample galaxies are listed in table 1.
%
% table 1, figure 1
%

Among the sample galaxies, 11 are found to have Keplerian rotation curves.
We use the term {`}Keplerian rotation curve{'} for a rotation curve which shows a significant drop of rotation velocity and can be fitted by a Keplerian, $V\sim R^{-1/2}$.
We did not include galaxies which show only a slight decrease of rotation velocity, such as discussed in Salucci and Frenk (1989).
Some examples of such rotation curves are shown in figure 1 with a Keplerian curve for a point mass as a reference.
The outermost parts of these rotation curves are approximated well by Keplerians.
We  comment on the properties of the Keplerian galaxies.
\\
{\bf NGC 891:} This Sb galaxy has been known to show a Keplerian drop for many years.
Sancisi and Allen (1979) first found a Keplerian drop on the southern part of the galaxy.
Although a counterpart on the north was not seen in their data, the VLA data by Rupen (1991) clearly show a drop on the north.
The rotation curve starts declining at the optical edge, and the velocity decreases by about 60 \vk.
\\
{\bf NGC 1365:} This is a grand-design barred spiral (SBb).
The {\sc Hi} rotation curve already starts declining inside of the optical disk (J\"ors\"ater, van Moorsel 1995).
The rotation velocity decreases by more than 50 \vk in total, and by 35 \vk in the Keplerian region.
\\
{\bf NGC 2683} The {\sc Hi} rotation curve of NGC 2683 was obtained by Casertano and van Gorkom (1991).
The rotation velocity decreases by 50 \vk outside the optical disk.
Although the rotation curve shows some irregularity, the drop in the rotation velocity is significant, and one may take it as being approximately Keplerian.
\\
{\bf NGC 3031} The {\sc Hi} in the outermost region of NGC 3031 is strongly distorted by the interaction with NGC 3034.
However, the {\sc Hi} velocity field is rather symmetric within the optical disk.
The rotation velocity is already decreasing in the optical disk (Rots, Shane 1975; Visser 1980).
\\
{\bf NGC 3521} The rotation curve of this Sb galaxy was obtained by Casertano and van Gorkom (1991).
The rotation velocity is almost constant at 200 \vk out to 8.5$h$ ($h$ is the disk scale length), and decreases by 45 \vk beyond it.
\\
{\bf NGC 4138} This Sa galaxy has a counter-rotating disk in the central region.
The {\sc Hi} rotation curve obtained by Jore et al.(1996) shows a significant drop by more than 100 \vk.
The rotation curve is well approximated by a Keplerian in the outer region.
However, it cannot be ruled out that the declining rotation curve may be caused by a very strong warp of the {\sc Hi} disk.
\\
{\bf NGC 4244} This is an edge-on Sc galaxy.
A Keplerian drop outside of the optical disk is seen on both sides after the flat rotation with 100 \vk (Olling 1996).
\\
{\bf NGC 4414} The {\sc Hi} rotation curve for this Sc galaxy was obtained by Braine et al.(1993).
The velocity decrease is evident by 70 \vk, starting at $4h$.
\\
{\bf NGC 5204} This is a dwarf galaxy with a maximum rotation velocity of 75 \vk.
Sicotte et al.(1997) have modeled the {\sc Hi} velocity field, taking the variation of the inclination into account, and derived the declining rotation curve in the outermost region.
They have constructed a mass model with the disk and the dark halo, and argued that a cutoff of the dark halo is necessary to account for the declining rotation curve.
\\
{\bf NGC 7793} This is an Sd galaxy with a moderate inclination.
Carignan and Puche (1990) have obtained a declining rotation curve by taking the inclination variation into account.
The rotation velocity drops from 110 \vk by about 20 \vk outside 4.5$h$.
Carignan and Puche (1990) argued that even a no-dark-halo model is possible for this galaxy.
\\
{\bf DDO 154} Carignan and Beaulieu (1989) have found a declining rotation curve in this gas-rich dwarf irregular galaxy.
The maximum rotation is 50 \vk, and the decline starts beyond 10$h$.
In spite of the Keplerian rotation curve, one needs a significant amount of dark matter to reproduce the rotation curve.
The mass model by Carignan and Beaulieu showed that 90$\%$ of the mass within the Keplerian region consists of dark matter.

%
% section 3
%
\section{Analysis and Results}

The uncertainty in the rotation velocity may be crucial for uncovering Keplerian rotation curves, as described in section 1.
In order to investigate this, we define the quantity $f$, which is the ratio of both sides in inequality (1),
\beq
f \equiv V(R_{\rm dec})\left[1-\left(\frac{R_{\rm dec}}{R_{\rm out}}\right)^{1/2}\right]/ 2 \sigma.
\eeq
When we have only a few measurement points between $R_{\rm dec}$ and $R_{\rm out}$, $f$ gives a criterion for discriminating a Keplerian rotation curve from flat rotation curves.
One requires $f\ge 1$ to detect a declining rotation curve.

We applied this criterion to non-Keplerian rotation curves.
Since we do not know $R_{\rm dec}$ for non-Keplerian rotation curves, we assumed $R_{\rm dec}=5h$ in this section.
This corresponds to a typical size of the optical disk (van der Kruit and Searle 1982; van der Kruit 1988).
This is also close to the value of $R_{\rm dec}$ for Keplerian rotation curves,
 since $R_{\rm dec}$ for the Keplerian rotation curves in the present sample ranges from $3h$ to $12h$, and its mean is $6.2h$.
If the rotation curve of a galaxy is not observed beyond $5h$, we use the rotation velocity at the observed outermost point, $V_{\rm out}$, as $V_{5h}$.

The velocity uncertainties are unknown for several galaxies.
For these galaxies, we give the following estimate.
The velocity uncertainty depends on the observational velocity resolution, the S/N ratio of the data, the uncertainty of the inclination angle, and so on.
The velocity uncertainty can be much less than the velocity resolution, if the S/N ratio is sufficiently high.
In the outermost region, however, the S/N ratio is just above the detection limit and the velocity profile could be fairly uncertain due to a possible decrease in the rotation velocity.
These facts make it difficult to achieve much a smaller uncertainty than the observational velocity resolution.
The ratio of the velocity uncertainties to the observational velocity resolution in table 1 is between 0.5 and 11, with a mean of 3.5 .
Therefore, the velocity uncertainty in the outermost region was assumed to be equal to $\sim 0.3 \Delta V/\sin i $, for galaxies for which the velocity uncertainty was not reported.

Figure 2 plots the value of $f$ versus $(R_{\rm out}-R_{\rm dec})/\Delta R$ for both Keplerian and non-Keplerian rotation curves.
%
% figure 2
%
$\Delta R$ is defined by $\Delta R \equiv D\surd \overline{a^2 + b^2}$, where $a$ and $b$ are the major and minor axis of the synthesized beam of the observation in radian, and $D$ is the distance to the galaxy.
The value of $(R_{\rm out}-R_{\rm dec})/\Delta R$ roughly corresponds to the number of independent measurements between $R_{\rm dec}$ and $R_{\rm out}$.
Figure 2 shows that the galaxies with Keplerian rotation curves indeed have an $f$ larger than unity, except for NGC 5204.
All but NGC 5204 have an $f$ larger than 2.5.
This indicates that an $f$ slightly larger than unity is not sufficient, and that more than $2$ is required for uncovering a Keplerian rotation curve.
A remarkable aspect in figure 2 is the clear split of Keplerian and non-Keplerian galaxies depending on $f$;
galaxies with a Keplerian rotation curve mostly lie above $f=2.5$, while many galaxies with non-Keplerian rotation curves lie below $f=1$.
The galaxies with $f\le 1$ have values of ($R_{\rm out}-R_{\rm dec}/\Delta R$) less than 2.
This confirms that there are only few independent measurements of the rotation velocities between $R_{\rm dec}$ and $R_{\rm out}$.
Any galaxies with rising rotation curves at $R_{\rm out}$ does not have a value of $f$ between 0 and 1.
These facts ensures that $f$ works as a good criterion.
The number of galaxies with $f$ greater than unity is 25, and 10 of them are galaxies with Keplerian rotation curves.
As far as galaxies with $f\ge 1$ are concerned, the fraction of Keplerian rotation curves is about two-fifths.
The results do not change drastically even when the assumed velocity uncertainties are reduced to half of what we have assumed above.
In that case, the number of galaxies with $f$ larger than unity is 27, and the fraction of the Keplerian rotation curve is still significant.
Therefore, Keplerian rotation curves are not rare.

On the other hand, 19 galaxies out of 45 were found to have $f$ smaller than unity.
A negative value of $f$ indicates that the {\sc Hi} observation does not extend beyond $5h$.
One cannot determine whether the rotation curves of these $f\le 1$ galaxies are flat beyond the optical disk.
Since a significant fraction of $f\ge 1$ galaxies have Keplerian rotation curves, there could be several Keplerian rotation curves in galaxies with $f\le 1$, which have not yet been discovered.

%
% section 4
%
\section{Discussion}
\subsection{Extent of Dark Halo}

While a large number of galaxies have $f$ smaller than 1, the non-Keplerian rotation curves with $f\ge 1$ provide direct evidence for extended dark halos outside of the optical disks.
How far do they extend ?
In order to constrain the minimum extent of the dark halos, one can use the criterion [equation (2)] with larger values of $R_{\rm dec}$: if $f$ becomes unity with a certain value of $R_{\rm dec}$, this radius gives the minimum extent of the dark halo that is consistent with the observed flat rotation curves within the uncertainty.
We calculated $f$ for non-Keplerian galaxies while varying the value of $R_{\rm dec}$ from $5h$, $6h$, ..., to $10h$.
%
% figure 3
%
Figure 3 shows the number distribution of $f \le 1$ galaxies and $f \ge 1$ galaxies out of 34 non-Keplerian galaxies for different values of $R_{\rm dec}$.
It was checked that galaxies with $f \le 1$ have only few points of independent measurements between $R_{\rm dec}$ and $R_{\rm out}$ in any case of $R_{\rm dec}$. 
Figure 3 demonstrates that the number of $f \ge 1$ galaxies decreases drastically with increasing the assumed value of $R_{\rm dec}$.
This indicates that the number of rotation curves that claim extended dark halos to large radii is quite small.
Only 4 rotation curves provide evidence for a dark halo extending to $10h$.
Therefore, rather small halos, which have cutoff radii within $10h$, are still consistent with most of the {\sc Hi} observations.
The existence of Keplerian rotation curves supports such small halos, if they reflect true mass truncations.

\subsection{Comparison with Other Studies}
Although the idea of a small halo conflicts with several studies of dark-matter halos which claim massive halos extending to several hundred kpc (e.g., Zaritsky, White 1994; Navarro et al.1996), there have been several indications of small halos other than rotation-curve studies.
Gottesman and Hunter (1982) analyzed the motion of satellite galaxies of NGC 3992, and argued that NGC 3992 is unlikely to have a halo with a mass of $\sim 10^{12}M_\odot$.
Dubinski et al.(1996) simulated the interacting galaxies, and concluded that galaxies with small halos, the rotation curve of which starts to decline within 10$h$, are favorable in order to reproduce the tidal tails of interacting galaxies.
Bland-Hawthorn et al.(1997) observed the ionized hydrogen in NGC 253 outside of the {\sc Hi} disk, and measured the rotation velocity where an {\sc Hi} observation could not reach.
Their result suggests that the rotation curve is declining beyond the {\sc Hi} disk, and that the halo is truncated.
All of these results are consistent with the idea of small halos.
One possibility to reconcile the discrepancy between small and massive halos is that the size and the mass of halos varies significantly from galaxy to galaxy.

On the other hand, the {\sc Hi} rotation curves, including Keplerian ones, could still be consistent with massive halos. 
One may reproduce a declining rotation curve if a specific mass distribution is assumed, as the velocity drop at the edge of the optical disk (e.g., Salucci, Frenk 1989).
Unfortunately, however, the extent of the {\sc Hi} rotation curve and its uncertainty are not sufficient to discriminate these two scenarios.
Extensive studies of the rotation velocities in the far outer region, where we hardly detected the {\sc Hi} gas, are necessary.
A large number of observations of ionized gas beyond the {\sc Hi} disk may lead to more decisive conclusions.
\vspace{1pc}\par

%
% acknowledgment
%
We thank the referee, P. Salucci, for his helpful comments.
M.H. also acknowledges the financial support from the Japan Society for the Promotion of Science.

%
% figure captions
%
\vspace{0.7cm}

\section*{Figure Captions}
\re {\bf Fig. 1.} Examples of declining rotation curves with eye-fit Keplerian rotation curves for point masses.
The data are taken from the references listed in table 1.

\re {\bf Fig. 2.} Plot of $f$ versus $(R_{\rm out}-R_{\rm dec})/\Delta R$.
The filled circles represent galaxies with a Keplerian rotation curve.
The crosses represent those with a non-Keplerian rotation curve.

\re {\bf Fig. 3.} Number distributions of non-Keplerian galaxies with $f\le 1$ and those with $f\ge 1$ with varying $R_{\rm dec}$.
The shadowed regions correspond to galaxies with $f\ge 1$.
%\clearpage

\section*{Notes for Table 1}

{\bf Columns :} (1) galaxy's name; (2) type (3) inclination (4) adopted distance in Mpc; (5) disk scale length in kpc; (6) rotation velocity at $5h$ in \vk; (7) radius at which rotation curve starts to decline (kpc); (8) rotation velocity at $R_{\rm dec}$ in \vk; (9) {\sc Hi} outermost radius in kpc; (10) rotation velocity at $R_{\rm out}$ in \vk, an asterisk denotes that the rotation curve shows a tendency of rising in the outermost region; (11) observational velocity resolution in \vk; (12) beam size of observation in arcsec; (13) rotation velocity uncertainty at $R_{\rm out}$ in \vk; (14) references :

 1 Puche et al.(1991); 2 Newton, Emerson (1977); 3 Kent (1987); 4 Carignan, Puche (1990b); 5 Carignan et al.(1991); 6 Newton (1980); 7 Rupen (1991); 8 van der Kruit, Searle (1982); 9 England (1989); 10 J\"or\"ater, van Moorsel (1995); 11 Broeils (1992); 12 Wevers et al.(1986); 13 Casertano, van Gorkom (1991); 14 Bosma (1981); 15 Meurer et al.(1996); 16 Rots, Shane (1975); 17 Visser (1980); 18 Jobin, Carignan (1990); 19 Begeman (1989); 20 Ball (1986); 21 Bottema (1996); 22 Jore et al.(1996); 23 Shostak, Rogstad (1973); 24 Olling (1996); 25 van Albada, Shane (1975); 26 Braine et al.(1993); 27 Guhathakurta et al.(1988); 28 Cayatte et al.(1990); 29 Kodaira et al.(1986); 30 Bottema et al.(1986); 31 Sicotte, Carignan (1997); 32 C\^ote et al.(1991); 33 Sancisi, van Albada (1987); 34 Carignan, Puche (1990a); 35 Martimbeau et al.(1994); 36 Carignan, Beaulieu (1989); 37 Lake et al.(1990); 38 van der Hulst et al.(1993); 39 Roelfsema, Allen (1985); 40 Kent (1986); 41 Cox et al.(1996); 42 de Blok et al.(1996).

\clearpage
%
% table 1
%
%\begin{table*}
\begin{center}
Table~1. Data for sample galaxies.
\vspace{6pt}
\begin{tabular}{cccccccccccccc}
\hline\hline
Galaxy & Type & $i$ & $D$ & $h$ & $V_{5h}$ & $R_{\rm dec}$ & $V_{\rm dec}$ & $R_{\rm out}$ & $V_{\rm out}$ & $\Delta V$ & $a\times b$ & $\sigma$ &Note \\
\hline
NGC  55 & SBm &79 &1.60&1.6& 85&---&---&10.2& 86&10.3&45 45&5.5&  1\\
NGC 224 & Sb  &77 &0.67&5.2&240&---&---& 30 &240&16.0&420 600&---& 2,3\\
NGC 247 & Sc  &75 &2.53&2.9&---&---&---& 10 &108&10.3&33 20&9.2&  4\\
NGC 253 & SBc &80 &2.58&2.4&---&---&---& 8.6&224*&20.6&68 42&20.8& 5\\
NGC 598 & Scd &54 &0.69&2.4&---&---&---&  6 &105*&16.0&90 180&2.7& 6\\
NGC 891 & Sb  &88 & 9.5&4.9&170& 15&220& 28 &160&20.7&40 40&---& 7,8\\
NGC 1300& SBbc&50 &17.1&7.0&---&---&---& 16 &150&25.2&20 20&40& 9\\
NGC 1365& SBb &55 &20.0&7.4&215& 24&260& 39 &190&20.8&12  6&---& 10\\
NGC 1560& Sd  &82 & 3.0&1.3& 75&---&---& 8.3& 80*& 8.2&13 14&2.3& 11\\
NGC 2403& Sc  &55 &3.25&2.1&130&---&---&19.5&130&16.5&45 45&---& 3, 12\\
NGC 2683& Sb  &70 & 5.1&1.2&200&  9&195& 18 &150&20.6&63 55& 15& 13\\
NGC 2841& Sb  &65 & 9.0&2.3&280&---&---& 36 &280&27.2&51 65& 10& 3, 14\\
NGC 2903& Sc  &60 & 6.1&1.9&200&---&---& 24 &190&33.1&33 33&---& 3, 12\\
NGC 2915& I   &60 & 5.3&0.7& 75&---&---& 15 & 92*& 3.3&45 45&1.3& 15\\
NGC 3031& Sab &59 &3.25&2.5&190&8.5&225&12.5&190&27.0&24 26&7.0& 3, 16, 17\\
NGC 3109& SBm &75 & 1.7&1.6& 66&---&---& 8.2& 66&10.3&36 27&3.2& 3, 18\\
NGC 3198& Sc  &70 & 9.2&2.5&150&---&---&  29&150&33.0&25 35&3.0& 3, 19\\
NGC 3359& SBc &51 &11.0&4.4&---&---&---&13.1&140&25.2&18 18& 21& 20\\
NGC 3521& Sbc &75 & 8.9&2.4&200&20.5&200&28.5&155&20.7&74 53&11.5& 13\\
NGC 4013& Sbc &90 &12.0&3.4&170&---&---& 22 &170&33.2&13 19& 10& 8, 21\\
NGC 4138& Sa  &52 &16.0&1.2&220&  7&200& 20 &120& 5.2&19 21&---& 22\\
NGC 4236& Sb  &75 &3.25&2.7&---&---&---& 11 & 85& 21&120 120&---& 3, 23\\
NGC 4244& Scd &85 & 5.0&2.6&100& 14&100& 22 & 82& 6.2&38 38&3.0& 8, 24\\
NGC 4258& Sbc &72 & 6.6&5.6&200&---&---& 30 &200& 27& 30 41&---& 3, 25\\
NGC 4414& Sc  &60 & 9.6&1.4&200&5.6&215& 13 &150& 33& 15 28&5.0& 26\\
NGC 4501& Sb  &56 &16.6&4.0&300&---&---& 18 &300&41.3&40 39&---& 27, 28, 29\\
NGC 4565& Sb  &85 &10.0&5.5&250&---&---& 32 &250&20.8&40 40&---& 7, 8\\
NGC 4654& Scd &49 &16.6&3.2&200&---&---& 16 &200*&25.0&48 43&---& 27, 28, 29\\
NGC 5023& Scd &90 & 8.0&2.0&---&---&---& 9.3& 80&33.3&14 21&---& 8, 30\\
NGC 5033& Sbc &60 &14.0&6.0&200&---&---& 36 &200&27.2&51 85&9.0& 3, 14\\
NGC 5055& Sbc &55 & 8.0&3.8&190&---&---&42.5&180&27.2&49 73&13.3& 3, 14\\
NGC 5204& Sm  &45 & 4.8&0.8& 75&8.7& 76&10.8& 68& 8.3&29 27&10.5& 31\\ 
NGC 5585& Sd  &52 & 6.2&1.4& 90&---&---& 9.6& 90&16.5&24 31& 2.7& 32\\
NGC 5907& Sc  &85 &11.0&5.7&220&---&---& 29 &220& 27 &60 60&---& 8, 33\\
NGC 7331& Sb  &70 &14.0&4.7&240&---&---& 30 &250&27.2&25 45&10.3& 3, 14\\
NGC 7793& Sd  &54 & 3.4&1.1&107&4.9&110& 7.4& 91&10.3&44 31& 6.1& 4\\
IC  2574& SBm &77 & 3.0&2.2&---&---&---& 8.1& 66*& 8.2&15 12& 15& 35\\
DDO 154 & Sm  &57 & 4.0&0.5& 37&5.8& 48& 7.5& 43&10.3&34 35&2.0& 36\\
DDO 170 & Im  &84 &14.6&1.7& 60&---&---&12.5& 66*&10.4&24 20&1.4& 37\\
UGC 0128& Sdm &57 & 60 &6.4&130&---&---& 42&130&21.2&25 23&---& 38\\
UGC 2885& Sc  &64 & 79 &12.7&300&---&---& 73&300&68.6&14 23&---& 39, 40\\
UGC 5750& SBdm&64 & 56 &3.3& 72&---&---& 22 & 75&21.2&24 22&---& 38\\
UGC 5999& Im  &55 & 45 &4.4&---&---&---& 15 &150*&21.1&24 21&---& 38\\
UGC 7170& Scd &90 &31.8&5.1&---&---&---&15.7&100&10.5&13 13&---& 41\\
F563-V1 & I   &60 & 38 &1.8& 24&---&---& 5.5& 24&20.6&13 14&---& 42\\
\hline
\end{tabular}
\end{center}

%\end{table*}
\clearpage

%
% references
%
\section*{References}

\re Ball R. 1986, ApJ 307, 453

\re Begeman K. G. 1989, A\&A 223, 47

\re Bland-Hawthorn J., Freeman K. C., Quinn P. J. 1997, ApJ in press (astro-ph/9706210)

\re Bosma A. 1981, AJ 86, 1791

\re Bottema R. 1996, A\&A 306, 345

\re Bottema R., Shostak G. S., van der Kruit P. C. 1986, A\&A 167, 34

\re Braine J., Combes F., van Driel W. 1993, A\&A 280, 451

\re Broeils A. H. 1992, A\&A 256, 19

\re Carignan C., Beaulieu S. 1989, ApJ 347, 760

\re Carignan C., Freeman K. C. 1985, ApJ 294, 494

\re Carignan C., Puche D. 1990, AJ 100, 394

\re Carignan C., Puche D. 1990, AJ 100, 641

\re Carignan C., Puche D., van Gorkom J. 1991, AJ 101, 456

\re Casertano S., van Gorkom J. H. 1991, AJ 101, 1231

\re Cayatte V., van Gorkom J. H., Balkowski C., Kotanyi C. 1990, AJ 100, 604

\re C\^ote S., Carignan C., Sancisi R. 1991, AJ 102, 904

\re Cox A. L., Sparke L. S., van Moorsel G., Shaw M. 1996, AJ 111, 1505

\re de Blok W. J. G., McGaugh S. S., van der Hulst J. M. 1996, MNRAS 283, 18

\re Dubinski J., Mihos J. C., Hernquist L. 1996, ApJ 462, 576

\re England M. N. 1989, ApJ 337, 191

\re Gottesman S. T., Hunter J. H. Jr. 1982, ApJ 260, 65

\re Guhathakurta P., van Gorkom J. H., Kotanyi C. G., Balkowski C. 1988, AJ 96, 851

\re Honma M., Sofue Y. 1996, PASJ 48, L103

\re Jobin M., Carignan C. 1990, AJ 100, 648

\re Jore K. P., Broeils A. H., Haynes M. P. 1996, AJ 112, 438

\re J\"ors\"ater S., van Moorsel G. A. 1995, AJ 110, 2037

\re Kent S. M. 1986, AJ 91, 1301

\re Kent S. M. 1987, AJ 93, 816

\re Kodaira K., Watanabe M., Okamura S. 1986, ApJS 62, 703

\re Lake G., Schommer R. A., van Gorkom J. H. 1990, AJ 99, 547

\re Martimbeau N., Carignan C., Roy J.-R. 1994, AJ 107, 543

\re Mathewson D. S., Ford V. L., Buchhorn M. 1992, ApJS 81, 413

\re Meurer G., Carignan C., Beauliel S. F., Freeman K. C. 1996, AJ 111, 1551

\re Navarro J. F., Frenk C., White S. D. M. 1996, ApJ 462, 563

\re Newton K. 1980, MNRAS 190, 689

\re Newton K., Emerson D. T. 1977, MNRAS 181, 573

\re Olling R. P. 1996, AJ 112, 457

\re Persic M., Salucci P. 1991, ApJ 368, 60

\re Persic M., Salucci P., Stel F. 1996, MNRAS 281, 27

\re Puche D., Carignan C., Wainscoat R. J. 1991, AJ 101, 447

\re Roelfsema P. R., Allen R. J. 1985, A\&A 146, 213

\re Rots A. H., Shane W. W. 1975, A\&A 45, 25

\re Rubin V. C., Burstein D., Ford W. K. Jr., Thonnard N. T. 1985, ApJ 289, 81

\re Rubin V. C., Thonnard N. T., Ford W. K. Jr. 1980, ApJ 238, 471 

\re Rupen M. P. 1991, AJ 102, 48

\re Salucci P., Frenk C. 1989, MNRAS 237, 247

\re Sancisi R., Allen R. J. 1979, A\&A 74, 73

\re Sancisi R., van Albada T. S. 1987, in Dark Matter in the Universe, IAU symposium No. 117, ed J. Kormendy, G. R. Knapp (Reidel, Dordrecht) p67

\re Shostak G. S., Rogstad D. H. 1973, A\&A 24, 405

\re Sicotte V., Carignan C. 1997, AJ 113, 609

\re Sofue Y. 1996, ApJ 458, 120

\re Sofue Y. 1997, PASJ 49, 17

\re Sofue Y., Honma M., Arimoto N. 1995, A\&A 296, 33

\re van Albada G. D., Shane W. W. 1975, A\&A 42, 433

\re van der Hulst J. M., Skillman E. D., Smith T. R., Bothun G. D., McGaugh S. S.,  de Blok W. J. G. 1993, AJ 106, 548

\re van der Kruit P. C. 1988, A\&A 192, 117

\re van der Kruit P. C., Searle L. 1982, A\&A 110, 61

\re Visser H. C. D. 1980, A\&A 88, 149

\re Wevers B. M. H., van der Kruit P. C., Allen R. J., 1986, A\&AS 66, 502

\re Zaritsky D., White S. D. M. 1994, ApJ 435, 599

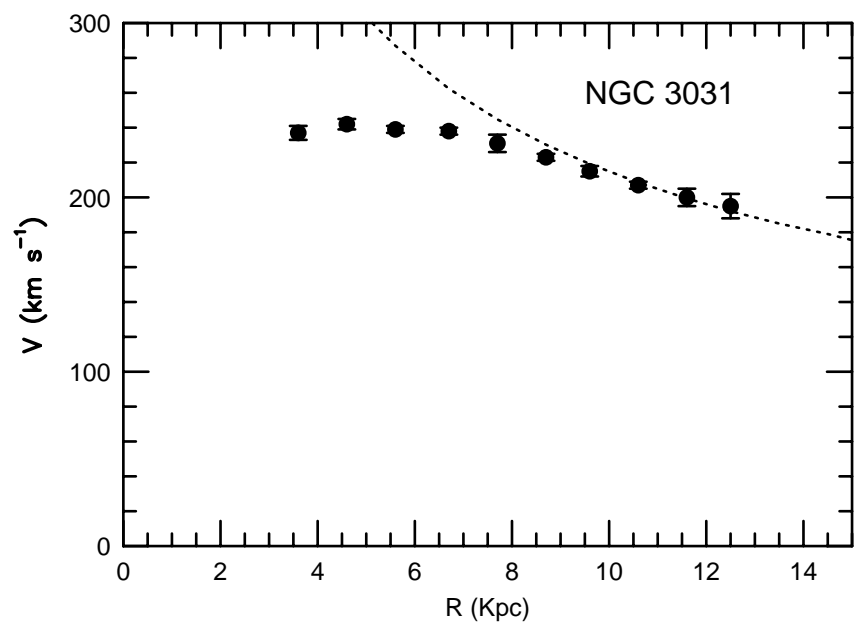
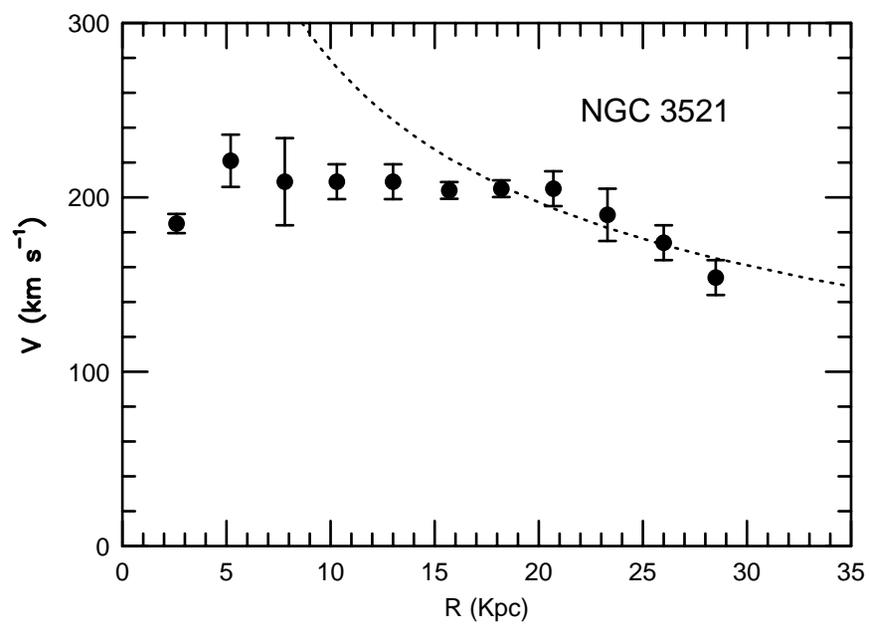
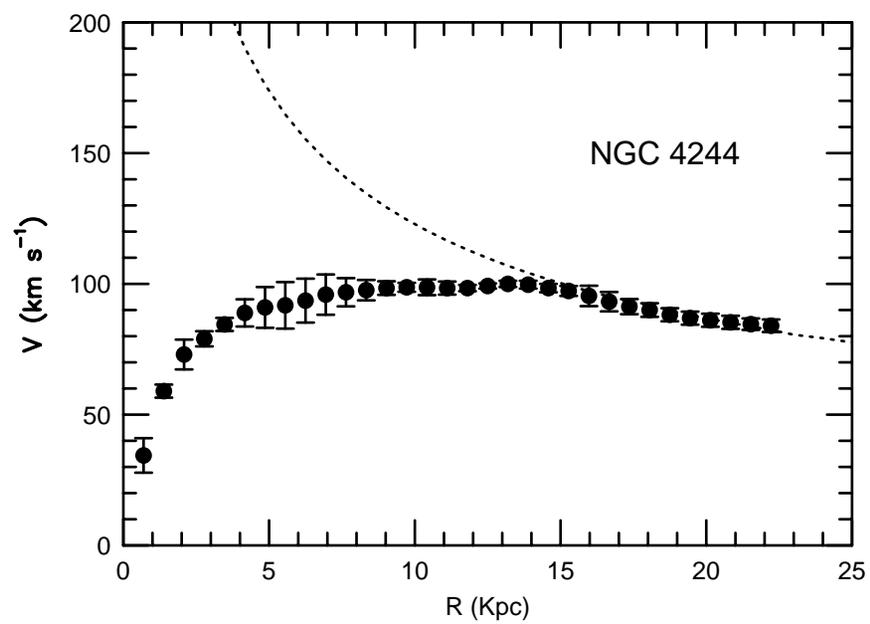
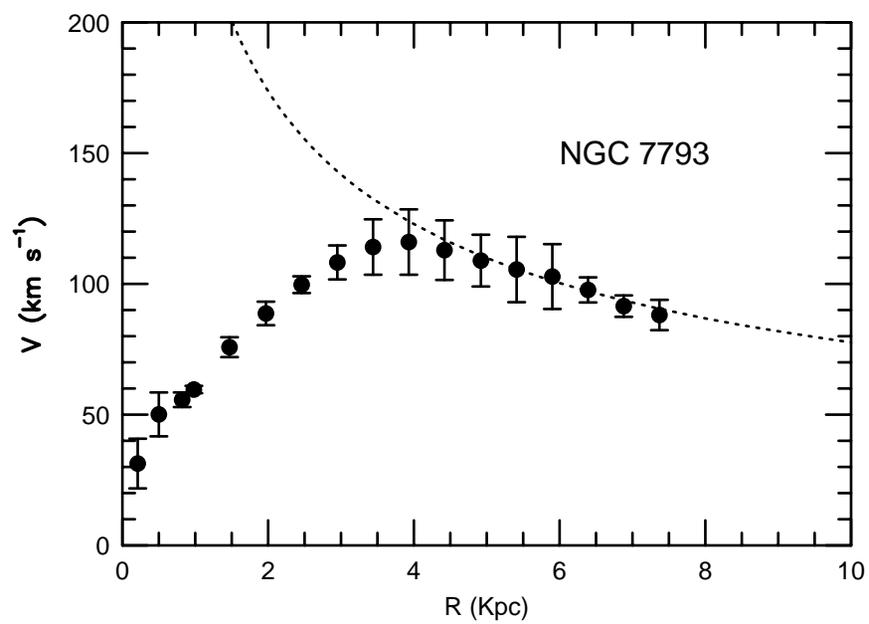
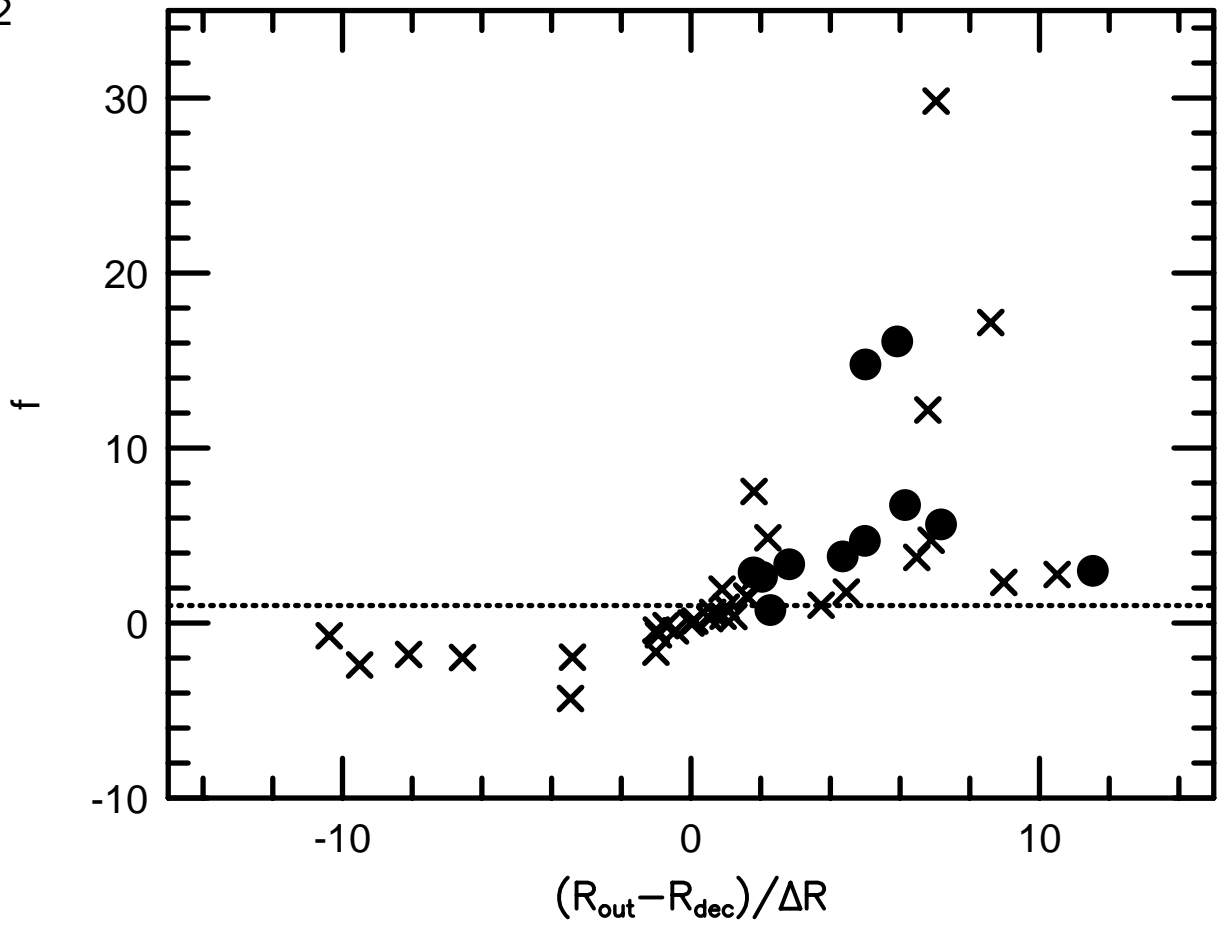
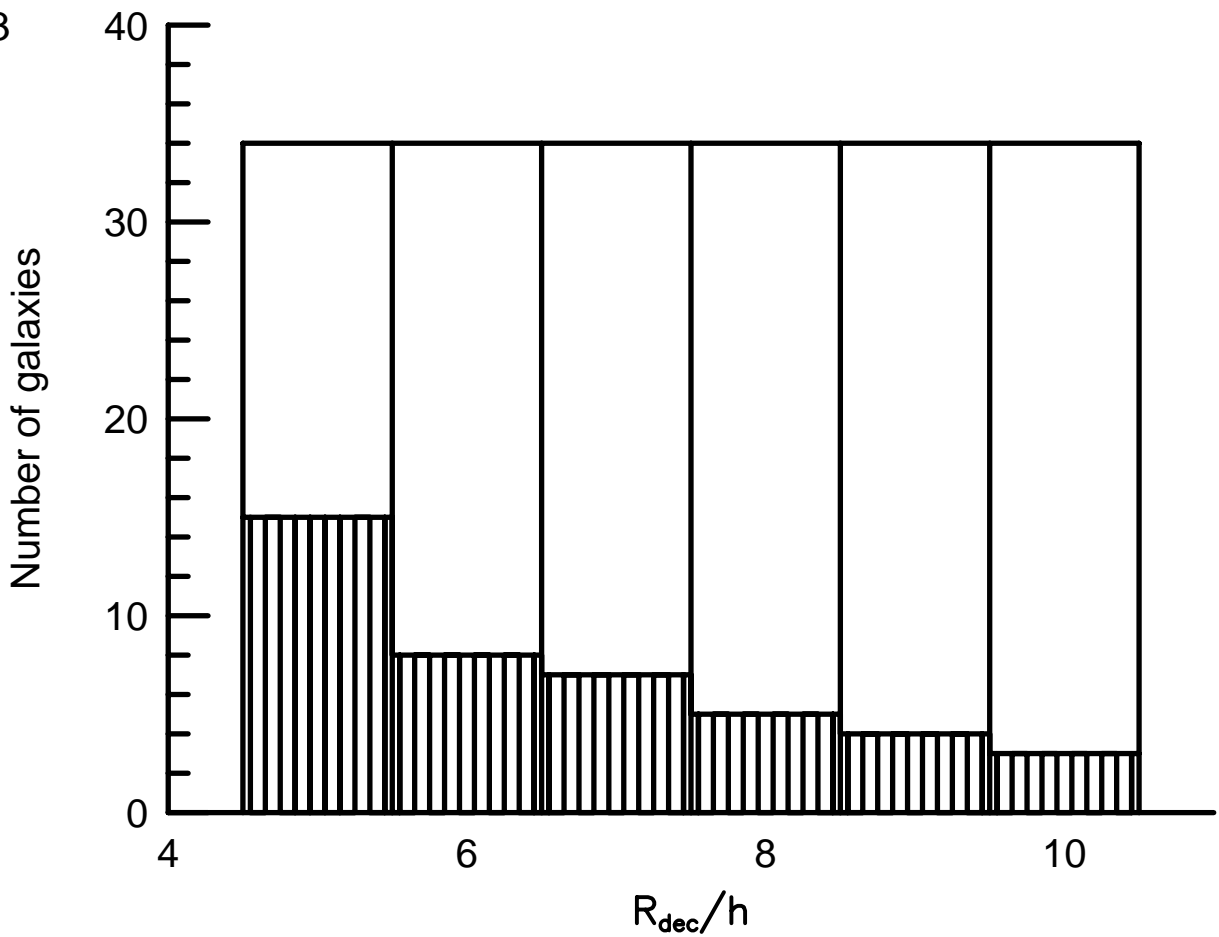
\end{document}